\documentclass[pre, twocolumn, showkeys, amsmath, amssymb]{revtex4}

\usepackage{graphicx}

\begin{document}

\title{When Does Division of Labor Lead to Increased System Output?}

\author{Emmanuel Tannenbaum}
\email{emanuelt@bgu.ac.il}
\affiliation{Department of Chemistry, Ben-Gurion University of the Negev,
Be'er-Sheva 84105, Israel}

\begin{abstract}

This paper develops a set of simplified dynamical models with which to explore
the conditions under which division of labor leads to optimized system output,
as measured by the rate of production of a given product.  We consider two
models:  In the first model, we consider the flow of some resource into
a compartment, and the conversion of this resource into some
product.  In the second model, we consider the resource-limited growth of
autoreplicating systems.  In this case, we divide the replication and metabolic tasks 
among different agents.  The general features that emerge from our models is that division 
of labor is favored when the resource to agent ratio is at intermediate values, and when the time 
cost associated with transporting intermediate products is small compared to characteristic process times.  
We discuss the results of this paper in the context of simulations with digital life.  We also argue that division 
of labor in the context of our replication model suggests an evolutionary basis for the emergence of the stem-cell-based 
tissue architecture in complex organisms.

\end{abstract}


\maketitle

\section{Introduction}

Division of labor is a ubiquitous phenomenon in biology.  In sufficiently complex multicellular organisms, 
various tasks necessary for organismal survival (metabolism, nutrient transport, motion, reproduction,
information processing, etc.) are performed by distinct parts of the organism \cite{ORGDIV1, ORGDIV2}.  Division of labor is
even possible in clonal populations of free-living single-celled organisms \cite{BACTDIV}.  At longer length 
scales, it is apparent that division of labor is a strong characteristic of community behavior
in various animals \cite{INSECTDIV1, INSECTDIV2}.  Human-built modern economies exhibit considerable
division of labor (indeed, much research into this phenomenon has been done by economists) \cite{ECONDIV1, ECONDIV2, ECONDIV3,
ECONDIV4, ECONDIV5, ECONDIV6, ECONDIV7, ECONDIV8}.

Selective pressure for the division of labor in a population of agents (cells, organisms, humans) arises because
specialization allows a given agent to optimize its performance of a relatively limited set of tasks.  Total 
system production of a population of differentiated agents can therefore be significantly greater 
than a comparable population of undifferentiated agents.

The question that arises then, is why is division of labor not always observed?  For example, while complex
multicellular organisms are certainly ubiquitous, approximately $ 80\% $ of the biomass of the planet is
in the form of bacteria.  While capable of exhibiting cooperative behavior, bacteria are, for the most part, 
free-living single-celled organisms.  Clearly then, there are regimes where differentiation is not desirable.  

As a general rule, the more complex the organism, the greater the selective pressure for differentiation
of system tasks.  This rule is admittedly somewhat circular, since a more complex organism will by definition
exhibit more specialization of the component agents.  So, to be more precise, the greater the number of
agents comprising a system, the greater the selective pressure for differentiation (even this formulation
has some ambiguity, because we can arbitrarily define any group of agents to comprise a system, no matter
how weak the inter-agent interactions.  Nevertheless, despite this ambiguity, we will proceed with this
initial ``working'' rule).

The origin of this rule comes from the observation that there is a cost associated with differentiation,
namely a time (and energy, though this will be ignored in this paper) cost associated with transporting 
intermediate products from one part of the system to another.  As system size grows, then presumably the density 
of agents grows (since the number of agents grows, and since we are grouping all the agents into one system, 
the inter-agent interactions are sufficiently strong, compared to some reference interaction, to warrant this grouping.  Note that increasing
the agent density is a simple way to do this, though highly interconnected systems may interact fairly strongly 
over relatively long distances.  The internet is a good example of this).  As the density of agents grows,
the characteristic time associated with transporting intermediates from one part of the system to another
decreases, and so the cost of differentiation decreases (in fairness, the idea of transport costs placing
a barrier to differentiation is not originally the author's.  In the context of firms, this idea has been
presented in the economics literature \cite{ECONDIV5}).  

In this paper, we develop two sets of models that capture the competition between the benefits of differentiation
and the time cost associated with differentiation.  In the first model, we consider the flow of some resource
into a compartment, and its conversion into some final product.  In the undifferentiated case, we assume that
there is a single agent capable of converting the resource into final product.  In the differentiated case,
we assume that the conversion of resource is accomplished in a two-step process, each of which is carried
out by distinct agents specialized for the separate tasks.

In the second model, we consider the flow of resource into a region containing replicating agents.  We assume
that the agents increase their volume so as to maintain a constant, pre-specified population density.  In the
undifferentiated case, we assume that a given agent can absorb the resource, and process it to produce a new
agent.  In the differentiated case, we assume a division of labor between replication and metabolism steps.
That is, we assume that a fraction of the agents are specialized so that they cannot replicate, but can only
process the resource into an intermediate form.  This metabolized resource is then processed by the replicators,
which produce new agents.  These daughter agents then undergo a differentiation step, where they can either
become replicators themselves, or metabolizers.

In both the compartment and the replicator models, the general features that emerge is that differentiation
is favored when population density is at intermediate levels with respect to resource numbers (when the population
density is low, then the undifferentiated pathways are favored, while when resources are highly limited, then
the difference between the undifferentiated and differentiated pathways disappears).  In the context of the 
replicator-metabolism model, we argue that this phenomenon suggests an evolutionary basis for the stem-cell-based 
tissue architecture in complex vertebrate organisms.

This paper is organized as follows:  In Section II, we develop and discuss the compartment model, involving the conversion
of some resource into a final product.  In Section III, we develop and discuss the replication-metabolism model.
In Section IV, we conclude with a summary of our results and plans for future research.

\section{Compartment Model}

\subsection{Definition of the model}

The compartment model is defined as follows:  Some resource, denoted $ R $, flows into a compartment of
volume $ V $ at a rate $ f_R $.  In the undifferentiated case, a single agent, denoted $ E $, processes
the resource to produce a final product, denoted $ P $ (the term $ E $ comes from chemistry, since the chemical
analogue of an agent is an \underline{E}nzyme catalyst).

In the differentiated case, the processing of $ R $ is accomplished by two separate agents, $ E_1 $ and $ E_2 $.
The agent $ E_1 $ first converts the resource into an intermediate product $ R^{*} $, and then the agent
$ E_2 $ converts $ R^{*} $ into $ P $.

It should be apparent that separating the tasks associated with converting $ R $ to $ P $ among
two different agents can only increase the total production rate of $ P $ if $ E_1 $ and $ E_2 $
can each perform their individual tasks better than $ E $.  Therefore, an implicit assumption
here is that, when an agent specializes, its ``native'' ability to perform a given task can
be made better than when an agent is unspecialized.

For a simple reason why this is true, let us imagine that $ E $, $ E_1 $, $ E_2 $ are enzymes,
i.e. protein catalysts, whose function is pre-determined by some amino acid sequence of length $ L $.
If the alphabet is of size $ S $, then there are $ S^L $ distinct sequences that can generate $ E $,
$ E_1 $, $ E_2 $.  Assuming that $ E_1 $ and $ E_2 $ are optimized for their particular functions,
we note that, in the absence of any additional information, the probability that $ E_1 $ and $ E_2 $
are the same is $ 1/S^{L} \rightarrow 0 $ as $ L \rightarrow \infty $.  Indeed, the average
Hamming distance (number of sites where two sequences differ) between any two sequences in the sequence
space is given by $ L (1 - 1/S) \rightarrow \infty $ as $ L \rightarrow \infty $.  

Therefore, it is highly likely that $ E $ is neither optimized for any of the tasks associated with
converting $ R $ to $ P $, but performs each task with some intermediate efficiency.

\subsection{Undifferentiated model}

In order to describe the processes governing the conversion of $ R $ to $ P $, we will adopt
the language and notation of chemical reaction kinetics.  This formalism is very convenient,
and is easily translatable into a system of ordinary differential equations.

For the undifferentiated model, we have,
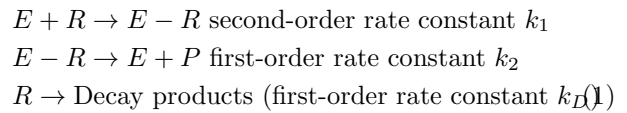
\begin{eqnarray}
&   &
E + R \rightarrow E-R \mbox{ second-order rate constant $ k_1 $}
\nonumber \\
&   &
E-R \rightarrow E + P \mbox{ first-order rate constant $ k_2 $}
\nonumber \\
&   &
R \rightarrow \mbox{Decay products (first-order rate constant $ k_D $)}
\end{eqnarray}
The first reaction refers to agent $ E $ grabbing the resource $ R $ (in chemistry, this
is referred to as the binding step).  At this point, the agent is denoted $ E-R $, to indicate
that it is bound to a resource particle.  In the second reaction, the agent processes the resource
to form the product $ P $, which it then releases.
 
The last reaction indicates that the resource $ R $ has a finite lifetime inside
the compartment, and decays with some first-order rate constant $ k_D $.  This assumption
ensures that the compartment cannot be filled with resource without limit.  The finite lifetime
can be due to back-diffusion of resource outside the compartment, or simply that the resource does
not last forever (some analogies include waiting time of customers at a restaurant before leaving without
being served, the characteristic time for a food product to spoil, or the diffusion of solute out of a cell).

If $ n_R $, $ n_E $, $ n_{ER} $ denote the number of particles of resource $ R $, unbound agents $ E $,
and agent-resource complexes $ E-R $, respectively, then we have,
\begin{eqnarray}
&   &
\frac{d n_R}{dt} = f_R - \frac{k_1}{V} n_E n_R - k_D n_R
\nonumber \\
&   &
\frac{d n_E}{dt} = -\frac{k_1}{V} n_E n_R + k_2 n_{ER}
\nonumber \\
&   &
\frac{d n_{ER}}{dt} = \frac{k_1}{V} n_E n_R - k_2 n_{ER}
\end{eqnarray}

If we define $ n = n_E + n_{ER} $, then note that $ d n/dt = d n_E/dt + d n_{ER}/dt = 0 $,
which implies that $ n $ is constant.  After some manipulation, we obtain that the steady-state
solution of this model is given by,
\begin{equation}
n_{R, ss} = \frac{k_2 n_{ER, ss}}{(k_1/V) (n - n_{ER, ss})}
\end{equation}
where $ n_{ER, ss} $ satisfies,
\begin{equation}
n_{ER, ss}^2 - (n + \frac{f_R}{k_2} + \frac{k_D}{(k_1/V)}) n_{ER, ss} + \frac{f_R}{k_2} n = 0
\end{equation}
so that,
\begin{equation}
n_{ER, ss} = \frac{1}{2}[(n + \frac{f_R}{k_2} + \frac{k_D}{(k_1/V)}) - \sqrt{(n + \frac{f_R}{k_2} + \frac{k_D}{(k_1/V)})^2 - 4 \frac{f_R}{k_2} n}] 
\end{equation}
We take the ``-'' root because it guarantees that $ n_{ER, ss} \leq n $ for all positive values of $ f_R $.

For small $ n $, a Taylor expansion of the quadratic to first order gives,
\begin{equation}
n_{ER, ss} = \frac{f_R/k_2}{f_R/k_2 + k_D/(k_1/V)} n \mbox{ (small $ n $)}
\end{equation}
while for large $ n $, Taylor expansion to first order with respect to the remaining terms gives,
\begin{equation}
n_{ER, ss} = \frac{f_R}{k_2} \mbox{ (large $ n $)}
\end{equation}

As a rough estimate of where the transition from the small $ n $ to the large $ n $ behavior occurs,
we can equate the two expressions and solve for $ n $.  The result is,
\begin{equation}
n_{trans, 1} = \frac{f_R}{k_2} + \frac{k_D}{(k_1/V)}
\end{equation}

\subsection{Differentiated model}

The conversion of resource $ R $ into $ P $ via the differentiated pathway occurs via 
the following sets of chemical reactions:
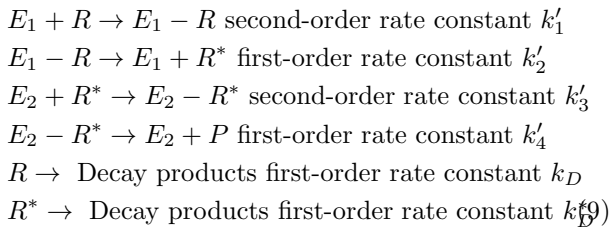
\begin{eqnarray}
&   &
E_1 + R \rightarrow E_1-R \mbox{ second-order rate constant $ k_1' $}
\nonumber \\
&   &
E_1-R \rightarrow E_1 + R^{*} \mbox{ first-order rate constant $ k_2' $}
\nonumber \\
&   &
E_2 + R^{*} \rightarrow E_2-R^{*} \mbox{ second-order rate constant $ k_3' $}
\nonumber \\
&   &
E_2-R^{*} \rightarrow E_2 + P \mbox{ first-order rate constant $ k_4' $}
\nonumber \\
&   &
R \rightarrow \mbox{ Decay products} \mbox{ first-order rate constant $ k_D $}
\nonumber \\
&   &
R^{*} \rightarrow \mbox{ Decay products} \mbox{ first-order rate constant $ k_D^{*} $}
\end{eqnarray}

Note that the intermediate product, $ R^{*} $, is also capable of decaying.  As we will see shortly,
it is the finite lifetime of the intermediate products that causes the undifferentiated pathway to
outperform the differentiated pathway at low agent numbers, and allows for a transition at higher
agent numbers, whereby the differentiated pathway overtakes the undifferentiated pathway.

In the context of our model, the direct interpretation of the decay term for $ R^{*} $ is that the
intermediate product has a finite lifetime, due either to diffusion out of the compartment or due to
decay into other compounds.  More generally, though, this term may refer to an aging cost, and 
therefore this model may be useful in understanding aspects of networked systems, whose function
does not necessarily depend on material transfers, but on information transfers.  

Information is transmitted between various parts of a system in order to effect system behavior, in response to the 
state of the system at the time of information transfer.  Therefore, there is a time limit during which the
information is relevant (because of the dynamic nature of the system and environment), which may be roughly modelled
by assuming that the information is ``lost'' via a first-order decay.

Defining particle and agent numbers analogously to the undifferentiated case, we obtain the system of
equations,
\begin{eqnarray}
&   &
\frac{d n_R}{dt} = f_R - \frac{k_1'}{V} n_{E_1} n_R - k_D n_R
\nonumber \\
&   &
\frac{d n_{R^*}}{dt} = k_2 n_{E_1 R} - \frac{k_3'}{V} n_{E_2} n_{R^*} - k_D^{*} n_{R^*}
\nonumber \\
&   &
\frac{d n_{E_1}}{dt} = -\frac{k_1'}{V} n_{E_1} n_R + k_2' n_{E_1 R}
\nonumber \\
&   &
\frac{d n_{E_1 R}}{dt} = \frac{k_1'}{V} n_{E_1} n_R - k_2' n_{E_1 R}
\nonumber \\
&   &
\frac{d n_{E_2}}{dt} = -\frac{k_3'}{V} n_{E_2} n_{R^{*}} + k_4' n_{E_2 R^{*}}
\nonumber \\
&   &
\frac{d n_{E_2 R^{*}}}{dt} = \frac{k_3'}{V} n_{E_2} n_{R^{*}} - k_4' n_{E_2 R^{*}}
\end{eqnarray}

If we define $ n_1 = n_{E_1} + n_{E_1 R} $ and $ n_2 = n_{E_2} + n_{E_2 R^{*}} $, then
note that $ d n_1/dt = d n_2/dt = 0 $, so that $ n_1 $ and $ n_2 $ are also constant.
Proceeding to solve for the steady-state of this model, we obtain,
\begin{widetext}
\begin{eqnarray}
&   &
n_{E_1 R, ss} = \frac{1}{2} [(n_1 + \frac{f_R}{k_2'} + \frac{k_D}{(k_1'/V)}) - 
\sqrt{(n_1 + \frac{f_R}{k_2'} + \frac{k_D}{(k_1'/V)})^2 - 4 \frac{f_R}{k_2'} n_1}]
\nonumber \\
&   &
n_{E_2 R^{*}, ss} = \frac{1}{2} [(n_2 + \frac{k_2 n_{E_1 R}}{k_4'} + \frac{k_D^{*}}{(k_1'/V)}) -
\sqrt{(n_2 + \frac{k_2' n_{E_1 R}}{k_4'} + \frac{k_D^{*}}{(k_3'/V)})^2 - 4 \frac{k_2' n_{E_1 R}}{k_4'} n_2}] 
\end{eqnarray}
\end{widetext}

Now, when $ n_1 $ and $ n_2 $ are small, it may be shown that to lowest non-vanishing order,
the steady state population of $ E_2-R^{*} $ is given by,
\begin{equation}
n_{E_2 R^{*}, ss} = \frac{(k_3'/V)}{k_D^{*}} \frac{k_2'}{k_4'} \frac{f_R/k_2'}{f_R/k_2' + k_D/(k_1'/V)} \alpha (1 - \alpha) n^2 \mbox{ (small $ n $)}
\end{equation}
where $ n \equiv n_1 + n_2 $, and $ \alpha \equiv n_1/n $, and $ 1 - \alpha = n_2/n $.

For large values of $ n $, we obtain that,
\begin{equation}
n_{E_2 R^{*}, ss} = \frac{f_R}{k_4'} \mbox{ (large $ n $)}
\end{equation}

As an estimate of where the transition between the small $ n $ and large $ n $ behavior occurs, we can equate the two
expressions and solve for $ n $.  We obtain,
\begin{equation}
n_{trans, 2} = \sqrt{\frac{1}{\alpha (1 - \alpha)} \frac{k_D^{*}}{(k_3'/V)} (\frac{f_R}{k_2'} + \frac{k_D}{(k_1'/V)})}
\end{equation}

\subsection{Comparison of undifferentiated and differentiated models}

The small $ n $ expression for the rate of production of final product in the undifferentiated case is,
\begin{equation}
k_2 n_{ER} = k_2 \frac{f_R/k_2}{f_R/k_2 + k_D/(k_1/V)} n
\end{equation}
while the small $ n $ expression for the rate of production of final product in the differentiated case is,
\begin{equation}
k_4' n_{E_2 R^{*}} = k_2' \frac{(k_3'/V)}{k_D^{*}} \frac{f_R/k_2'}{f_R/k_2' + k_D/(k_1'/V)} \alpha (1 - \alpha) n^2
\end{equation}

Note then that for sufficiently small $ n $, the undifferentiated production pathway produces final product
more quickly than the differentiated pathway.  However, because the rate of production of final product 
for the undifferentiated pathway initially increases linearly with $ n $, while the rate of production 
of final product for the differentiated pathway increases quadratically, it is possible that the differentiated
pathway eventually overtakes the undifferentiated pathway.  The critical $ n $ where this is estimated to 
occur, denoted $ n_{equal} $, may be estimated by equating the two expressions.  The final result is,
\begin{equation}
n_{equal} = \frac{1}{\alpha (1 - \alpha)} \frac{k_D^{*}}{(k_3'/V)} \frac{f_R/k_2' + k_D/(k_1'/V)}{f_R/k_2 + k_D/(k_1/V)}
\end{equation}

Now, for $ n_{equal} $ to be meaningful, it must occur in a regime where the rate expressions used to obtain it are valid.
Therefore, we want $ n_{equal} < n_{trans, 1}, n_{trans, 2} $.  However, we can make an even stronger statement.  If
$ n_{equal} $ does indeed refer to a point beyond which the differentiated pathway overtakes the undifferentiated pathway,
then we should have $ n_{equal} < n_{trans, 2} < n_{trans, 1} $.  It is possible to show that,
\begin{widetext}
\begin{equation}
\frac{n_{trans, 2}}{n_{equal}} = \frac{n_{trans, 1}}{n_{trans, 2}} = \sqrt{\alpha (1 - \alpha) \frac{(k_3'/V)}{k_D^{*}}
\frac{1}{f_R/k_2' + k_D/(k_1'/V)}} (\frac{f_R}{k_2} + \frac{k_D}{(k_1/V)})
\end{equation}
\end{widetext}
and so, our inequality is equivalent to the condition that,
\begin{equation}
\frac{k_D^{*}}{(k_3'/V)} < \alpha (1 - \alpha) (\frac{f_R}{k_2} + \frac{k_D}{(k_1/V)}) \frac{f_R/k_2 + k_D/(k_1/V)}{f_R/k_2' + k_D/(k_1'/V)}
\end{equation}
which implies that
\begin{equation}
n_{equal} < \frac{f_R}{k_2} + \frac{k_D}{(k_1/V)}
\end{equation}

Figures 4 and 5 (of the version submitted to {\it The Journal of Theoretical Biology}) show comparisons of the production rates of final product for the 
undifferentiated and differentiated pathways.  In Figure 4, the parameters are chosen so that the differentiated pathway eventually overtakes the 
undifferentiated pathway, while in Figure 5, the parameters are chosen so that this is not the case.  Note, however, that even in Figure 4, although the
differentiated pathway overtakes the undifferentiated pathway, once $ n $ becomes very large, the undifferentiated pathway
again overtakes the differentiated pathway.

This behavior can be explained as follows:  When $ n $ is very small, the rate at which the intermediate product is ``grabbed''
by $ E_2 $ is small compared to the decay rate, so that much intermediate product is lost.  In this regime, the undifferentiated
pathway is optimal, for, although $ E $ may be less efficient than either $ E_1 $ or $ E_2 $ at their respective tasks, the overall
production rate of $ P $ is not reduced by the loss of intermediates.

Now, as $ n $ increases, the rate of loss of intermediates decreases to an extent such that the increased efficiency associated
with differentiation causes the differentiated pathway to overtake the undifferentiated pathway.  However, once $ n $ increases 
even further, then there is sufficient quantity of agents in both the undifferentiated and differentiated pathways to process
all of the incoming resource, with minimal loss due to decay.  At this point, because the production rate of $ P $ has become
resource limited, the efficiency advantage of the differentiated pathway is considerably reduced, such that the slight cost
associated with intermediate decay becomes sufficient to cause the undifferentiated pathway to overtake the differentiated pathway.
However, this effect is a small one, since, once $ n $ is very large, both the undifferentiated and differentiated pathways perform
similarly.

\subsection{When can a differentiated pathway outperform an undifferentiated pathway?}

The analysis of the previous section deserves further scrutiny, in order to better understand
the circumstances under which differentiation can lead to improved system performance.  

At low agent numbers, the decay of the product intermediates leads to a quadratic increase in 
system output, and so the undifferentiated pathway outperforms the differentiated pathway.  
At some point, however, the number of agents is sufficiently large that the decay of both resource
and intermediates is minimal, so that it is possible for the differentiated pathway to overtake
the undifferentiated pathway.  However, this is only possible if the differentiated pathway is
more efficient than the undifferentiated pathway.  

To quantify this notion, assume that $ n $ is at some intermediate value, such that $ k_D $
and $ k_D^{*} $ may be effectively taken to be $ 0 $.  In this regime, it is possible to
show that,
\begin{equation}
k_4' n_{E_2 R^{*}, ss} = \min \{k_2' \alpha n, k_4' (1 - \alpha) n\}
\end{equation}
Essentially, if $ k_2' \alpha n > k_4' (1 - \alpha) n $, then the first set of agents are
capable of producing intermediate at a rate greater than the second set of agents are capable
of processing it, so that the second reaction step is rate limiting.  If $ k_2' \alpha n <
k_4' (1 - \alpha) n $, then the first reaction step is rate limiting.

Note then that if one of the reactions is rate limiting, we can adjust the agent fractions
to increase the rate of the rate limiting reaction, and thereby increase the overall production
rate of $ P $.  Therefore, the maximal production rate of $ P $ is achieved when
$ k_2' \alpha n = k_4' (1 - \alpha) n \Rightarrow \alpha_{optimal} = k_4'/(k_2' + k_4') $,
so that the maximal production rate of $ P $ is given by,
\begin{equation}
(k_4' n_{E_2 R^{*}, ss})_{max} = \frac{k_2' k_4'}{k_2' + k_4'} n
\end{equation}
For the undifferentiated case, the analogous expression is $ k_2 n $, and so we expect that
the differentiated pathway can only overtake the undifferentiated pathway when,
\begin{eqnarray}
&   &
\frac{k_2' k_4'}{k_2' + k_4'} > k_2 
\nonumber \\
&   &
\Rightarrow 
\frac{1}{k_2'} + \frac{1}{k_4'} < \frac{1}{k_2}
\end{eqnarray}

Intuitively, this condition makes sense, since $ 1/k_2 $ is the characteristic time it takes
agent $ E $ to convert $ R $ to $ P $, while $ 1/k_2' $ and $ 1/k_4' $ are the characteristic
times for agents $ E_1 $ and $ E_2 $ to perform their respective tasks.  Therefore, differentiation
can only overtake nondifferentiation if the characteristic time for the completion of a set of tasks
is shorter for the differentiated pathway than it is for the undifferentiated pathway. 

If $ 1/k_2' + 1/k_4' > 1/k_2 $, it is in principle possible for the differentiated pathway to 
overtake the undifferentiated pathway if $ k_1' $ is sufficiently greater than $ k_1 $, and
if $ k_3' $ is sufficiently large compared to $ k_D^{*} $.  Basically, the differentiated
agents are not more efficient at actually processing the resource, but they are more efficient
at grabbing them, which can give the differentiated pathway an advantage.  However, in contrast
to the case where $ 1/k_2' + 1/k_4' < 1/k_2 $, this advantage is only temporary, because once
the agent number becomes sufficiently large, the characteristic time to grab resource becomes
very small.

As a final note, because the condition for differentiation to outperform nondifferentiation 
at larger agent numbers is $ 1/k_2' + 1/k_4' < 1/k_2 $, while the agent number $ n_{equal} $
where differentiation overtakes nondifferentiation does not depend on $ k_4' $, it should be
apparent that our criterion for $ n_{equal} $ could be inaccurate in actually predicting the
location of the cross-over.  This is because $ n_{equal} $ is based on the small $ n $ region,
where the rate of production of $ P $ for the differentiated pathway increases quadratically.
This is the regime where the production rate of $ P $ is limiting by intermediate resource 
decay.

However, as $ 1/k_2' + 1/k_4' $ increases, we expect $ n_{equal} $ to become a better predictor
of the crossover point, since the decay of the intermediate resource becomes a comparatively
greater factor in dictating the performance of the differentiated pathway.

In any event, though, the expression for $ n_{equal} $ and the condition for the existence
of a cross-over are useful, for they indicate that the larger the value of $ f_R $, the 
larger the value of $ k_D^{*} $ that is possible for a cross-over to still occur.  In particular,
it suggests that, as long as $ 1/k_2' + 1/k_4' < 1/k_2 $, then by making $ f_R $ sufficiently
large for a given $ k_D^{*} $, we will eventually obtain that the differentiated pathway
will outperform the undifferentiated pathway at sufficiently high agent numbers.  This is indeed
what is observed numerically.

\section{Replication-Metabolism Model}

In this section, we turn our attention to the replication-metabolism model, where a population
of agents processes an external resource for the purposes of producing more agents.

\subsection{Definition of the model}

We consider a population of replicating agents, relying on the supply of some resource,
denoted $ R $.  We assume that the resource is supplied to the population at a rate of
$ f_R $ per unit volume, and that, as the population grows, the volume expands in such
a way as to maintain a constant population density $ \rho $.

In the undifferentiated model, a single agent, denoted $ E $, processes the resource
$ R $ and replicates.  In the differentiated model, an agent $ E_1 $ ``metabolizes'' the
resource to some intermediate $ R^{*} $, and then another agent, denoted $ E_2 $, processes
the intermediate and reproduces.  However, the $ E_2 $ agents are responsible for
supplying both metabolizers and replicators.  Therefore, $ E_2 $ produces a ``blank'' agent,
denoted $ E $, which then specializes and becomes either $ E_1 $ and $ E_2 $.

\subsection{Undifferentiated model}

The reactions defining the undifferentiated model are,
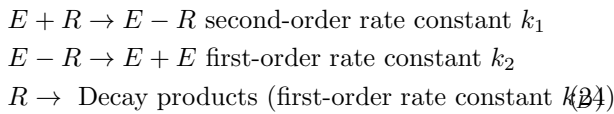
\begin{eqnarray}
&   &
E + R \rightarrow E-R \mbox{ second-order rate constant $ k_1 $}
\nonumber \\
&   &
E-R \rightarrow E + E \mbox{ first-order rate constant $ k_2 $}
\nonumber \\
&   &
R \rightarrow \mbox{ Decay products (first-order rate constant $ k_D $)}
\end{eqnarray}

In terms of population numbers, the dynamical equations for $ n_E $, $ n_{ER} $ and $ n_R $ are,
\begin{eqnarray}
&   &
\frac{d n_E}{dt} = -\frac{k_1}{V} n_E n_R + 2 k_2 n_{ER}
\nonumber \\
&   &
\frac{d n_{ER}}{dt} = \frac{k_1}{V} n_E n_R - k_2 n_{ER}
\nonumber \\
&   &
\frac{d n_R}{dt} = f_R V - \frac{k_1}{V} n_E n_R - k_D n_R
\end{eqnarray}
Therefore, defining $ n = n_E + n_{ER} $, we have,
\begin{equation}
\frac{d n}{dt} = k_2 n_{ER}
\end{equation}

Since the population density is $ \rho $, this implies that,
\begin{equation}
\frac{d V}{dt} = \frac{1}{\rho} \frac{d n}{dt} = k_2 V x_{ER}
\end{equation}
where $ x_E \equiv n_E/n $, $ x_{ER} \equiv n_{ER}/n $.

Now, the concentration $ c_R $ of the resource $ R $ is given
by the relation $ n_R = c_R V $, which implies that
\begin{eqnarray}
\frac{d c_R}{dt} 
& = & 
\frac{1}{V} (\frac{d n_R}{dt} - c_R \frac{d V}{dt})
\nonumber \\
& = & 
f_R - (k_1 \rho x_E + k_2 x_{ER} + k_D) c_R
\end{eqnarray}

Putting everything together, we obtain, finally, the
system of equations,
\begin{eqnarray}
&   &
\frac{1}{n} \frac{d n}{dt} = k_2 x_{ER}
\nonumber \\
&   &
\frac{d x_E}{dt} = -k_1 c_R x_E + 2 k_2 x_{ER} - k_2 x_{ER} x_E
\nonumber \\
&   &
\frac{d x_{ER}}{dt} = k_1 c_R x_E - k_2 x_{ER} - k_2 x_{ER}^2
\nonumber \\
&   &
\frac{d c_R}{dt} = f_R - c_R (k_1 \rho x_E + k_2 x_{ER} + k_D)
\end{eqnarray}

We can determine the steady-state behavior of the model by setting the left-hand-side
of the above system of equations to $ 0 $.  When $ \rho \rightarrow 0 $, the
steady-state solution is characterized by,
\begin{equation}
c_{R, ss} = \frac{f_R}{k_D + k_2 x_{ER, ss}} \mbox{ ($ \rho \rightarrow 0 $)}
\end{equation}
where $ x_{ER, ss} $ is the solution to the cubic,
\begin{equation}
0 = x^3 + (1 + \frac{k_D}{k_2}) x^2 + (\frac{k_1 f_R}{k_2^2} + \frac{k_D}{k_2}) x - \frac{k_1 f_R}{k_2^2}
\mbox{ ($ \rho \rightarrow 0 $)}
\end{equation}
Note that when $ f_R = 0 $, we obtain $ x_{ER, ss} = 0 $.  Therefore, differentiating the cubic with respect
to $ x $ gives,
\begin{equation}
(\frac{d x_{ER, ss}}{d f_R})_{f_R = 0} = \frac{k_1}{k_2 k_D}
\end{equation}
and so we have,
\begin{equation}
x_{ER, ss} = \frac{k_1}{k_2 k_D} f_R \mbox{ (small $ f_R $, $ \rho \rightarrow 0 $)}
\end{equation}

When $ f_R $ is large, we get $ x_{ER, ss} \rightarrow 1 $, so that
\begin{equation}
x_{ER, ss} = 1 \mbox{ ($ f_R \rightarrow \infty $, $ \rho \rightarrow 0 $)}
\end{equation}

Equating the small $ f_R $ and large $ f_R $ expressions, we obtain that the transition from small 
$ f_R $ to large $ f_R $ behavior is approximated by,
\begin{equation}
f_{R, trans, 1}(\rho = 0) = k_2 \frac{k_D}{k_1}
\end{equation}

Now, when $ \rho $ is large, then the steady-state expression for $ c_R $ is approximated
by,
\begin{equation}
0 = \frac{d c_R}{dt} = f_R - c_R k_1 \rho x_E \rightarrow k_1 c_R x_E = \frac{f_R}{\rho}
\end{equation}
and so,
\begin{equation}
0 = \frac{f_R}{\rho} - k_2 x_{ER, ss} - k_2 x_{ER, ss}^2
\end{equation}
from which it follows that,
\begin{equation}
x_{ER, ss} = \frac{1}{2}[-1 + \sqrt{1 + 4 \frac{f_R}{k_2 \rho}}]
\end{equation}
Since $ \rho $ is large, we will approximate this expression further, by taking the
first-order expansion in $ f_R/(k_2 \rho) $, giving,
\begin{equation}
x_{ER, ss} = \frac{f_R}{k_2 \rho}
\end{equation}

We can estimate the cross-over from small $ \rho $ to large $ \rho $ behavior by equating the
two expressions.  We have two estimates, one for small $ f_R $, and one for large $ f_R $.
We obtain,
\begin{eqnarray}
&   &
\rho_{trans, 1}^{-} = \frac{k_D}{k_1} \mbox{ ($ f_R < k_2 \frac{k_D}{k_1} $)}
\nonumber \\
&   &
\rho_{trans, 1}^{+} = \frac{f_R}{k_2} \mbox{ ($ f_R > k_2 \frac{k_D}{k_1} $)}
\end{eqnarray}

\subsection{Differentiated model}

The reactions defining the differentiated model are,
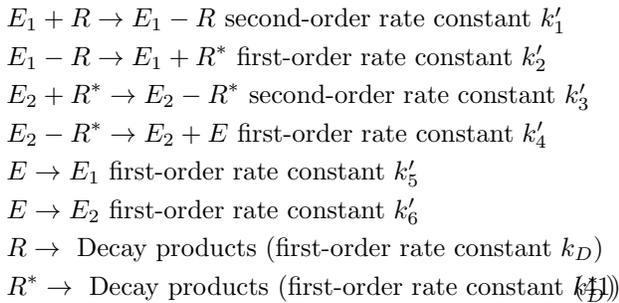
\begin{eqnarray}
&   &
E_1 + R \rightarrow E_1-R \mbox{ second-order rate constant $ k_1' $}
\nonumber \\
&   &
E_1-R \rightarrow E_1 + R^{*} \mbox{ first-order rate constant $ k_2' $}
\nonumber \\
&   &
E_2 + R^{*} \rightarrow E_2-R^{*} \mbox{ second-order rate constant $ k_3' $}
\nonumber \\
&   &
E_2-R^{*} \rightarrow E_2 + E \mbox{ first-order rate constant $ k_4' $}
\nonumber \\
&   &
E \rightarrow E_1 \mbox{ first-order rate constant $ k_5' $}
\nonumber \\
&   &
E \rightarrow E_2 \mbox{ first-order rate constant $ k_6' $}
\nonumber \\
&   &
R \rightarrow \mbox{ Decay products (first-order rate constant $ k_D $)}
\nonumber \\
&   &
R^{*} \rightarrow \mbox{ Decay products (first-order rate constant $ k_D^{*} $)} 
\end{eqnarray}

Following a procedure similar to the one carried out for the undifferentiated model, we obtain the
system of equations,
\begin{eqnarray}
&   &
\frac{1}{n} \frac{d n}{dt} = k_4' x_{E_2 R^{*}}
\nonumber \\
&   &
\frac{d x_{E_1}}{dt} = -k_1' c_R x_{E_1} + k_2' x_{E_1 R} + k_5' x_E - k_4' x_{E_2 R^{*}} x_{E_1}
\nonumber \\
&   &
\frac{d x_{E_1 R}}{dt} = k_1' c_R x_{E_1} - k_2' x_{E_1 R} - k_4' x_{E_2 R^{*}} x_{E_1 R}
\nonumber \\
&   &
\frac{d x_{E_2}}{dt} = -k_3' c_{R^{*}} x_{E_2} + k_4' x_{E_2 R^{*}} + k_6' x_E - k_4' x_{E_2 R^{*}} x_{E_2}
\nonumber \\
&   &
\frac{d x_{E_2 R^{*}}}{dt} = k_3' c_{R^{*}} x_{E_2} - k_4' x_{E_2 R^{*}} - k_4' x_{E_2 R^{*}}^2
\nonumber \\
&   &
\frac{d x_E}{dt} = k_4' x_{E_2 R^{*}} - (k_5' + k_6') x_E - k_4' x_{E_2 R^{*}} x_E
\nonumber \\
&   &
\frac{d c_R}{dt} = f_R - (k_1' \rho x_{E_1} + k_D) c_R - k_4' x_{E_2 R^{*}} c_{R^{*}}
\nonumber \\
&   &
\frac{d c_{R^{*}}}{dt} = \rho (k_2' x_{E_1 R} - k_3' c_{R^{*}} x_{E_2}) - (k_D^{*} + k_4' x_{E_2 R^{*}}) c_{R^{*}}
\nonumber \\
\end{eqnarray}

Now, defining $ \tilde{x}_{E_1} = x_{E_1} + x_{E_1 R} $, and $ \tilde{x}_{E_2} = 
x_{E_2} + x_{E_2 R^{*}} $, we obtain,
\begin{eqnarray}
&   &
\frac{d \tilde{x}_{E_1}}{dt} = k_5' x_E - k_4' x_{E_2 R^{*}} \tilde{x}_{E_1}
\nonumber \\
&   &
\frac{d \tilde{x}_{E_2}}{dt} = k_6' x_E - k_4' x_{E_2 R^{*}} \tilde{x}_{E_2}
\end{eqnarray}

Therefore, at steady-state we have,
\begin{equation}
\frac{\tilde{x}_{E_1, ss}}{\tilde{x}_{E_2, ss}} = \frac{k_5'}{k_6'}
\end{equation}
and so, using the relation $ \tilde{x}_{E_1} + \tilde{x}_{E_2} + x_E = 1 $ we obtain,
\begin{eqnarray}
&   &
\tilde{x}_{E_1, ss} = \frac{k_5'}{k_5' + k_6'} (1 - x_{E, ss})
\nonumber \\
&   &
\tilde{x}_{E_2, ss} = \frac{k_6'}{k_5' + k_6'} (1 - x_{E, ss})
\end{eqnarray}

If we let $ k_5', k_6' \rightarrow \infty $ such that $ k_5'/k_6' $ remains constant,
then it should be clear that $ x_{E, ss} \rightarrow 0 $.  Intuitively, $ E $
differentiates to either $ E_1 $ or $ E_2 $ as soon as it is produced, so it does not
build up in the system.  The ratio between $ k_5' $ and $ k_6' $ then dictates
the fraction of $ E_1 $ and $ E_2 $ in the system (allowing $ k_5', k_6' \rightarrow \infty $
essentially amounts to assuming that the differentiation time is zero.  This is of course
not true, and future research will need to incorporate positive differentiation times).

Defining $ \alpha = k_5'/(k_5' + k_6') $, we then have $ \tilde{x}_{E_1, ss} = \alpha $,
and $ \tilde{x}_{E_2, ss} = 1 - \alpha $.  Therefore, to characterize the system at
steady-state, we need to solve four equations, giving the steady-state conditions
for $ x_{E_1 R} $, $ x_{E_2 R^{*}} $, $ c_R $, and $ c_{R^{*}} $, respectively.
The equations are,
\begin{eqnarray}
&   &
0 = k_1' c_R (\alpha - x_{E_1 R}) - k_2' x_{E_1 R} - k_4' x_{E_2 R^{*}} x_{E_1 R}
\nonumber \\
&   &
0 = k_3' c_{R^{*}} (1 - \alpha - x_{E_2 R^{*}}) - k_4' x_{E_2 R^{*}} - k_4' x_{E_2 R^{*}}^2
\nonumber \\
&   &
0 = f_R - c_R (k_D + k_1' \rho (\alpha - x_{E_1 R})) - k_4' x_{E_2 R^{*}} c_R
\nonumber \\
&   &
0 = \rho (k_2' x_{E_1 R} - k_3' c_{R^{*}} (1 - \alpha - x_{E_2 R^{*}})) 
\nonumber \\
&   &
- k_D^{*} c_{R^{*}} - k_4' c_{R^{*}} x_{E_2 R^{*}}
\end{eqnarray}
As with the undifferentiated case, we study the behavior of this system of equations in both the small and large
$ \rho $ limits.  

When $ \rho = 0 $, we have $ c_{R^{*}, ss} = 0 \Rightarrow x_{E_2 R^{*}, ss} = 0 \Rightarrow
c_{R, ss} = f_R/k_D \Rightarrow x_{E_1 R, ss} = (k_1' f_R \alpha/k_D)/(k_2' + k_1' f_R/k_D) $.

Differentiating the steady-state equations with respect to $ \rho $, and evaluating at $ \rho = 0 $,
gives,
\begin{eqnarray}
&   &
(\frac{d c_{R^{*}, ss}}{d \rho})_{\rho = 0} = \frac{k_2'}{k_D^{*}} (x_{E_1 R})_{\rho = 0}
\nonumber \\
&   &
(\frac{d x_{E_2 R^{*}, ss}}{d \rho})_{\rho = 0} = \frac{k_3'}{k_4'} (1 - \alpha) (\frac{d c_{R^{*}, ss}}{d \rho})_{\rho = 0}
\end{eqnarray}
and so, for small $ \rho $, we have,
\begin{equation}
x_{E_2 R^{*}, ss} = \frac{k_2' k_3'}{k_4' k_D^{*}} \frac{\frac{k_1' f_R}{k_D}}{k_2' + \frac{k_1' f_R}{k_D}} \alpha (1 - \alpha) \rho \mbox{ (small $ \rho $)}
\end{equation} 

Now for large $ \rho $, our steady-state equations may be reduced to,
\begin{eqnarray}
&   &
0 = k_1' c_R (\alpha - x_{E_1 R}) - k_2' x_{E_1 R} - k_4' x_{E_2 R^{*}} x_{E_1 R}
\nonumber \\
&   &
0 = k_3' c_{R^{*}} (1 - \alpha - x_{E_2 R^{*}}) - k_4' x_{E_2 R^{*}} - k_4' x_{E_2 R^{*}}^2
\nonumber \\
&   &
0 = f_R - k_1' c_R \rho (\alpha - x_{E_1 R})
\nonumber \\
&   &
0 = k_2' x_{E_1 R} - k_3' c_{R^{*}} (1 - \alpha - x_{E_2 R^{*}})
\end{eqnarray}

The third equation gives $ k_1' c_R (\alpha - x_{E_1 R}) = f_R/\rho $, which may be substituted into
the first equation to give,
\begin{equation}
0 = \frac{f_R}{\rho} - x_{E_1 R} (k_2' + k_4' x_{E_2 R^{*}})
\end{equation}
Solving for $ x_{E_1 R} $ in terms of $ x_{E_2 R^{*}} $, and plugging the resulting
expression into the fourth steady-state equation gives, after some manipulation,
that $ x_{E_2 R^{*}, ss} $ is the solution of the cubic,
\begin{equation}
0 = x_{E_2 R^{*}, ss}^3 + (1 + \frac{k_2'}{k_4'}) x_{E_2 R^{*}, ss}^2 + \frac{k_2'}{k_4'} x_{E_2 R^{*}, ss} - \frac{k_2' f_R}{k_4'^2 \rho}
\end{equation}

Now, when $ f_R = 0 $, we obtain $ x_{E_2 R^{*}, ss} = 0 $.  From this it is possible
to show that,
\begin{equation}
(\frac{d x_{E_2 R^{*}, ss}}{d f_R})_{f_R = 0} = \frac{1}{\rho k_4'}
\end{equation}
and so,
\begin{equation}
x_{E_2 R^{*}, ss} = \frac{f_R}{k_4' \rho} \mbox{ (large $ \rho $)}
\end{equation}

As with the undifferentiated case, the transition from small $ \rho $ to 
large $ \rho $ behavior may be estimated by equating the two expressions and
solving for $ \rho $.  The result is,
\begin{equation}
\rho_{trans, 2} = \sqrt{\frac{1}{\alpha (1 - \alpha)} \frac{k_D k_D^{*}}{k_3'}(\frac{1}{k_1'} + \frac{f_R}{k_2' k_D})}
\end{equation}

\subsection{Comparison of undifferentiated and differentiated models}

As a function of $ f_R $, we wish to determine if, as $ \rho $ increases, the average growth
rate of the differentiated population overtakes the growth rate of the undifferentiated population.
If this does indeed happen, then there exists a $ \rho $, denoted $ \rho_{equal} $, at which 
the two growth rates are equal.

We first consider the regime $ f_R < k_2 (k_D/k_1) $.  This is the small $ f_R $ regime of
the undifferentiated population.  In this regime, the transition from low $ \rho $ to large
$ \rho $ behavior occurs at $ \rho_{trans, 1} = k_D/k_1 $.  For the differentiated pathway,
we have $ \rho_{trans, 2} = \sqrt{\frac{1}{\alpha (1 - \alpha)} \frac{k_D^{*}}{k_1' k_3'}(\frac{k_4'}{k_2'} f_R + k_D)} $.

Now, we have four possibilities:  (1) $ \rho_{equal} < \rho_{trans, 1}, \rho_{trans, 2} $.
(2) $ \rho_{trans, 2} < \rho_{equal} < \rho_{trans, 1} $.  (3) $ \rho_{trans, 1} < \rho_{equal}
< \rho_{trans, 2} $.  (4) $ \rho_{equal} > \rho_{trans, 1}, \rho_{trans, 2} $.

We can immediately eliminate Cases (2), (3), and (4) as possibilities.  For Case (4), we get an
undifferentiated rate of $ k_2 f_R/(k_2 \rho) = f_R/\rho $, and a differentiated rate
of $ k_4' f_R/(k_4' \rho) = f_R/\rho $, and so the two rates are equal.  For Case (2),
we get an undifferentiated rate of $ k_1 f_R/k_D $, and a differentiated rate of $ f_R/\rho $,
so equating gives $ \rho_{equal} = k_D/k_1 = \rho_{trans, 1} $.  Therefore, Case (2) is
essentially a limiting case of Case (4), and can also eliminated.  For Case (3), we get
an undifferentiated rate of $ f_R/\rho $, and a differentiated rate of 
$ (k_2' k_3'/k_D^{*}) (k_1' f_R/k_D)/(k_2' + k_1' f_R/k_D) \alpha (1 - \alpha) \rho $, so
equating gives $ \rho_{equal} = \rho_{trans, 2} $.  Therefore, Case (3) is essentially a limiting
case of Case (1), and can also be eliminated.

For Case (1), we have,
\begin{equation}
\rho_{equal} = \frac{1}{\alpha (1 - \alpha)} \frac{k_1 k_D^{*}}{k_3'} (\frac{1}{k_1'} + \frac{f_R}{k_2' k_D})
\end{equation}

Now, we can show that,
\begin{equation}
\frac{\rho_{equal}}{\rho_{trans, 2}} = \frac{\rho_{trans, 2}}{\rho_{trans, 1}} = k_1 \sqrt{\frac{1}{\alpha (1 - \alpha)}
\frac{k_D^{*}}{k_3' k_D} (\frac{1}{k_1'} + \frac{f_R}{k_2' k_D})}
\end{equation}
and so, in order for $ \rho_{equal} < \rho_{trans, 1}, \rho_{trans, 2} $, then we must have,
\begin{equation}
f_R < k_D \frac{k_2}{k_1} \frac{k_2'}{k_2} [\alpha (1 - \alpha) \frac{k_3' k_D}{k_1 k_D^{*}} - \frac{k_1}{k_1'}]
\end{equation}

We now consider the case where $ f_R > k_D \frac{k_2}{k_1} $.  This is the large $ f_R $ regime of the
undifferentiated population.  Following a similar procedure to the one carried out for the small $ f_R $ 
regime, we can show that the only possible crossover occurs in the small $ \rho $ regimes for both
the undifferentiated and differentiated cases.  In this regime, we obtain,
\begin{equation}
\rho_{equal} = \frac{k_2}{f_R} \frac{1}{\alpha (1 - \alpha)} \frac{k_D k_D^{*}}{k_3'} (\frac{1}{k_1'} + \frac{f_R}{k_2' k_D})
\end{equation}

We can show that,
\begin{equation}
\frac{\rho_{equal}}{\rho_{trans, 2}} = \frac{\rho_{trans, 2}}{\rho_{trans, 1}} = frac{k_2}{f_R} 
\sqrt{\frac{1}{\alpha (1 - \alpha)} \frac{k_D k_D^{*}}{k_3'} (\frac{1}{k_1'} + \frac{f_R}{k_2' k_D})}
\end{equation}
and so, in order for $ \rho_{equal} < \rho_{trans, 1}, \rho_{trans, 2} $, we must have,
\begin{equation}
\frac{k_D^{*}}{k_3'} < \alpha (1 - \alpha) \frac{(f_R/k_2)^2}{k_D/k_1' + f_R/k_2'}
\end{equation}

In Figure 9, we show a high-$ f_R $ plot where the differentiated growth rate overtakes the
undifferentiated growth rate.  In Figure 10, we show a high-$ f_R $ plot where the undifferentiated
growth rate stays above the differentiated rate at all values of $ \rho $ (these figures are included
in the version submitted to {\it The Journal of Theoretical Biology}).

\subsection{When can a differentiated population outreplicate an undifferentiated population?}

We can subject our replication-metabolism model to a similar analysis to the one applied
to the compartment model.  First of all, as with the compartment model, we expect that
the differentiated pathway can only overtake the undifferentiated pathway, and then
maintain a higher replication rate if $ 1/k_2' + 1/k_4' < 1/k_2 $.  Again, this condition
simply states that the total characteristic time associated with converting resource
into a new agent in the differentiated case is less than the total characteristic time
in the undifferentiated case.  The assumption is that decay costs are negligible, as well
as time costs associated with grabbing resource and intermediates.

An interesting behavior that occurs with the replication-metabolism model is the
different dependence on $ f_R $ that the transition population density $ \rho_{equal} $ has
in the low-$ f_R $ regime and the high-$ f_R $ regime.

In the high-$ f_R $ regime, $ \rho_{equal} $ has a weak dependence on $ f_R $, though it
does decrease as $ f_R $ increases.  This makes sense, for in the high-$ f_R $ regime, 
the growth rate of the undifferentiated population is limited by the rate at which
the complex $ E-R $ produces new agents.  As $ f_R $ increases, the cost associated
with the decay of the intermediate resource $ R^{*} $ decreases, so that the differentiated
pathway overtakes the undifferentiated pathway sooner.

In the low-$ f_R $ regime, $ \rho_{equal} $ increases linearly with $ f_R $, so that, as $ f_R $ 
increases in this regime, the differentiated pathway overtakes the undifferentiated pathway only
at higher values of $ \rho $ (if it overtakes at all).  The reason for this behavior is that at low $ f_R $, the growth
rate of the undifferentiated population is resource limited, so that increasing $ f_R $ actually
increases the growth rate.  The effect of this is to push to higher values of $ \rho $ the point
at which the differentiated agents outreplicate the undifferentiated agents.

What is interesting with these patterns of behavior is that they indicate opposite criteria
for when a cooperative replicative strategy is favored, depending on the availability of
resource:  When resources are plentiful, then increasing the resource favors
a differentiated replication strategy.  However, when resource are limited, then {\it decreasing} the
resource favors a differentiated replication strategy.

In this vein, it is interesting to note that complex multicellular life is only possible in relatively
resource-rich environments.  On the other hand, organisms such as the cellular slime mold ({\it Dictyostelium
discoideum}) transition from a single-celled to a multi-celled life cycle when starved.  While we have
already postulated one possible reason for this behavior in terms of minimizing overall reproductive
costs \cite{MULTICELL}, the behavior indicated in our model may provide another, complementary explanation
for the selective advantage for this phenomenon.

\section{Conclusions and Future Research}

This paper developed two models with which to compare the performance of undifferentiated and differentiated
pathways.  The first model considered the flow of resource into a compartment filled with a fixed number of
agents, whose collective task is to convert the resource into some final product.  The second model considered
the replication rate of a collection of agents, driven by an externally supplied resource.

By assuming that the resource, and even more importantly, that reaction intermediates, have a finite lifetime,
we were able to show that undifferentiated pathways are favored at low agent numbers and/or densities, while
differentiated pathways are favored at higher agent numbers and/or densities.  An equivalent way of stating
this is that differentiation is favored when resources are limited, where resource limitation is measured by
the ratio of available resource to agents.  

Some interesting results that emerged from our studies was that, although limited resources favor differentiation
(as measured by the resource-agent ratio), for a given set of system parameters, differentiation will be more
likely to overtake nondifferentiation at higher population size and/or density if the amount of available resource
is increased (although the actual cross-over location will increase as well).  The central reason for this is
that the relative decay costs associated with differentiation are decreased as resource is increased.

In the context of the replication-metabolism model, we should note that when resources are plentiful, differentiation
is favored at lower population densities as the resource flow is increased, while when resources are limited,
differentiation is favored at lower population densities as the resource flow is decreased.  Regarding the former
observation, it should be noted that it has been shown that diversity of replicative strategies
is favored at intermediate levels of resources \cite{DIFFWILKE}.  In digital life simulations, they showed that
the number of distinct replicating computer programs was maximized at intermediate resource availability, a
result consistent with what is observed ecologically.  We claim that the results of this paper are consistent
with these observations.

Regarding the latter observation, we pointed out in the previous section that this behavior is possibly
consistent with the behavior of organisms such as the cellular slime mold, which transition from a
single-celled to a multi-celled life form when starved.

We also posit that the results of the replication-metabolism model suggest a possible evolutionary basis
for the stem-cell-based tissue architecture in complex multicellular organisms.  Essentially, as population
density increases, and therefore as the resource-to-agent ratio decreases, it becomes more efficient for
some cells to exclusively focus on replicating and renewing the population, while other cells engage in
specialized functions necessary for organismal survival.  

Of course, our replication-metabolism model is not quite the same as a stem-cell-based tissue architecture.
First of all, the stem-cell and tissue cell population does not collectively grow.  Rather, the stem cells
periodically divide in order to replace dead tissue cells.  Therefore, the stem-cell-based tissue architecture
is a kind of hybrid between our compartment model and our replication-metabolism model.

Secondly, our replication-metabolism model assumes that there is a single differentiation step, while in reality
a differentiating tissue cell undergoes several divisions and differentiation steps before becoming a mature
tissue cells.

Finally, our replication-metabolism model assumed that differentiation was instantaneous.  In reality,
differentiation takes time, and this time cost will affect whether differentiation can overtake non-differentiation,
and, if so, will likely delay the critical population density where this happens.

Despite these shortcomings, we believe that the models developed here could be used as the basis for more sophisticated
models that could produce, via an optimization criterion, the stem-cell-based tissue architecture observed in complex
multicellular organisms.  This is a subject we leave for future work.

\begin{acknowledgments}

This research was supported by the Israel Science Foundation (Alon Fellowship).

\end{acknowledgments}

\end{document}